\documentclass[prl,twocolumn,showpacs]{revtex4}
\usepackage{dcolumn}
\usepackage{bm}
\usepackage{graphicx}

\begin{document}
\title{The song of the dunes as a self-synchronized instrument}

\author{S. Douady$^{\star}$, A. Manning$^{\star}$, P. Hersen$^{\star,\S}$, H. Elbelrhiti$^{\dag,\ddag}$, S. Proti\`{e}re$^{\star}$, A. Daerr$^{\dag}$, B.
Kabbachi$^{\ddag}$\\}

\affiliation{$^{\star}$ Mati\`{e}re et Syst\`{e}mes Complexes, Unit\'{e} Mixte de Recherche du Centre National de Recherche Scientifique \&
Universit\'{e} Paris 7.\\}

\affiliation{$^{\S}$ Bauer Laboratory, Center For Genomic Research, Harvard University, Cambridge, USA.\\}

\affiliation{$^{\dag}$ Physique \& M\'{e}canique des Milieux H\'{e}t\'{e}rog\`{e}nes, Ecole Sup\'{e}rieure de Physique-Chimie Industrielle,
Unit\'{e} Mixte de Recherche. du Centre National de Recherche Scientifique, 4 rue Vauquelin, 75005 Paris, France.\\}

\affiliation{$^{\ddag}$ GEOenvironement des Milieux ARIDes, Geology Department, Universit\'{e} Ibn Zohr, BP 28/5, Cité Dakhla, 80000 Agadir,
Morocco.}

\date{\today}

\begin{abstract}
Since Marco Polo \cite{M1298} it has been known that some sand dunes have the peculiar ability of emitting a loud sound with a well-defined
frequency, sometimes for several minutes. The origin of this sustained sound has remained mysterious, partly because of its rarity in nature.
It has been recognized that the sound is not due to the air flow around the dunes but to the motion of an avalanche, and not to an acoustic
excitation of the grains but to their relative motion. By comparing several singing dunes and two controlled experiments, one in the
laboratory and one in the field, we finally demonstrate here that the frequency of the sound is the frequency of the relative motion of the
sand grains. The sound is produced because some moving grains synchronize their motions. The existence of a velocity threshold in both
experiments further shows that this synchronization comes from an acoustic resonance within the flowing layer: if the layer is large enough
it creates a resonance cavity in which grains self-synchronize. Sound files are provided as supplementary materials.
\end{abstract}
%
%

\maketitle
%

In musical instruments, sustained sound is obtained through the coupling of excitation and resonance. The excitation is a more or less
periodic instability, like the rubbing stick-slip instability of the bow of a violin or the Von-Karmann whistling instability of a flute.
This excitation is coupled with a resonance (the string for the violin, the air volume for the flute). The coupling results in the adaptation
of the instability frequency to the one fixed by the resonance. Does the sound emitted by the sand dunes originate from a similar mechanism?
It has long been recognized that the song of dunes comes from the flowing motion of sand grains in an avalanche
\cite{D60,C23,LCCC76,NPB97,SBN97,CWG93}. Direct observations, as we have done in Morocco, Chile, and China, show that this sound does not
originate from stick-slip motion of blocks of grain (as the bow of a string instrument), because it is produced only by dry grains flowing
freely. Neither does it correspond to a resonance inside the dune (as in a wind instrument), because the same frequency has been measured at
different locations on a dune and, also, in the same field in dunes of different sizes.\\
\begin{table}
\begin{flushleft}
\footnotesize{
\begin{tabular}{lccc}
Location & Grains& Predicted & Measured\\
&size ($\mu m$)& frequency ($Hz$)& frequency ($Hz$)\\
\tableline\\
Ghord Lahmar &$160$    &$100$    & $105 \pm 10$\\
(Tarfaya, Morocco)&&&\\
Mar de Dunas &$210$    &$87$     & $90 \pm 10$\\
(Copiapo, Chile) &&&\\
Cerro Bramador&$270$    &$77$     & $75 \pm 10$\\
(Copiapo, Chile) &&&\\
Sand Mountain    &$340$    &$68$     & $63 \pm 5$\\
(Nevada, USA)* &&&\\
\tableline \tableline
\end{tabular}}
\caption{\footnotesize{Sound Frequency of Singing Dunes. Comparison of the frequency predicted from the grains size with the measured
frequency. *From \cite{LCCC76}; others are from our own measurements \cite{S01,S02,S03}. El Cerro Bramador sound properties were first
reported by Darwin \cite{D60}.}}
\end{flushleft}
\end{table}

\begin{figure}[h]
\centering
\includegraphics[width=80mm]{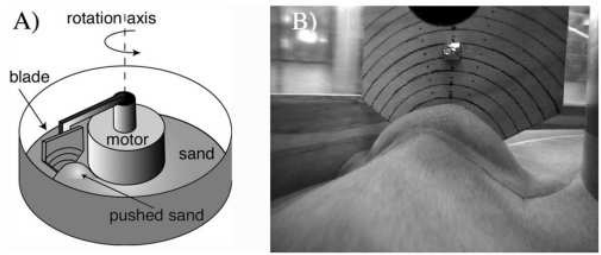}
\caption{\footnotesize{A: Sketch of laboratory experiment. Channel is of $1\, m$ diameter and $25\, cm$ width. B: Picture of flat pushing
blade in channel, taken by camera moving with blade. Black spot on top is microphone (also rotating). Circles on blade, separated by $1\,
cm$, allow direct measurement of height of pushed mass during motion}}\label{fig1}
\end{figure}
%

\begin{figure}[h]
\centering
\includegraphics[width=80mm]{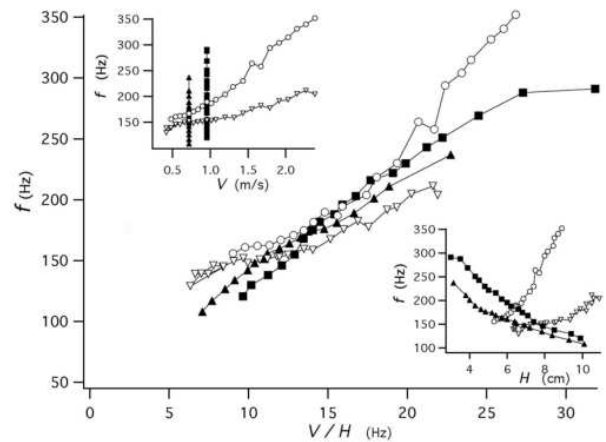}
\caption{\footnotesize{Frequency emitted by pushed (sheared) sand, measured in laboratory experiment, as a function of two laboratory control
parameters, height of mass of sand, $H$, and velocity of pushing blade, $V$. Four series of measurement are shown, two at constant pushing
velocity (black squares and upward triangles), two at constant plugging depth (gray open circles and downward triangles). Two insets show
that frequency depends on both parameters. In upper left, frequency varies with plugging depth, even at fixed velocity, and in lower right,
two curves have opposite variations with pushed height (two gray curves are not vertical as pushed height increases with velocity, even at a
constant plugging depth). Main plot shows that four curves collapse on a single one when plotted as a function of ratio of velocity with
height, which is mean shear rate to which mass of sand is submitted. Frequency observed is ten times mean shear, meaning that height of real
shear zone, between pushed mass and fixed bed, is on the order of $10\%$ of pushed sand height.}}\label{fig2}
\end{figure}

The first significant observation is that the frequency measured depends only on the size of the grains, each dune field having a
characteristic grain size and frequency. Study of the motion of the grains in an avalanche shows how grains have to pass periodically over
each other, dilating locally, and then hitting lower grains \cite{QADD00}. This particular motion explains that a grain flows with a constant
velocity and an avalanche with a linear velocity profile in depth \cite{R03}. The corresponding constant velocity gradient is precisely the
average shock frequency, $f$, identical throughout the avalanche depth. Experiments \cite{GDR04} and theory \cite{QADD00} give $f = 0.4
\sqrt{g/d}$, proportional to $\sqrt{g/d}$ as first proposed by \cite{B66}. This diameter to frequency relation is in accordance with what has
been measured (see Table 1) \cite{S01,S02,S03}. Even if there is a slight deviation, as noticed by \cite{LC94}, it is noticeable that the
prediction is within the error bars. In granular avalanches, this frequency is fixed by gravity, but pushing such "musical sand" with the
hand, a plate, or the legs can create different notes \cite{BJ85,L36} from $25$ to $250\, Hz$ in Morocco \cite{S04,S05} (cf. supplementary
recordings \cite{S01,S02,S03,S04,S05,S06,S07,S08,S09,S10,S11,S12} to listen to the sound of dunes). To reproduce this effect in the
laboratory, a blade is plugged at different depths into a prepared crest of singing sand (from Ghord Lahmar, Tarfaya, Morocco) and is pushed
by a motor at different velocities (see Figure~\ref{fig1}). This experiment allows an independent control of both the shearing velocity and
the mass of sheared sand, making it possible to obtain sustained sounds of constant and well-defined frequencies \cite{S06,S07}. This
demonstrates that the booming sound can be reproduced in a controlled way in the laboratory, and consequently that the dune itself is not
needed to produce the sound. The results in Figure~\ref{fig2} show that it is neither the velocity nor the mass that controls the frequency,
but the shear applied to the grains. This
demonstrates the hypothesis of \cite{PT22}, and is conclusive proof that what is being heard is the relative motion of the grains.\\

%


Singing dunes usually present well-sorted and rounded grains \cite{LCCC76,RV83}. This means that they can all share the same average motion
in the avalanche. Each grain, passing over lower grains and hitting them, creates a local sound wave with the associated dilatation,
compression, and shocks. All these events are desynchronized a priori in the flowing layer. All the pressure fluctuations then mainly cancel
out, leading to a light, high-frequency rustling, as is normally heard in sand avalanches \cite{S08}. So, how can a low-frequency sound be
produced? The only possibility is that a given number of grains start to move coherently. Then the whole flowing layer will move up and down,
and its surface will directly emit a pressure wave in the air, like a loudspeaker (or the belly of a violin). The surface motion amplitude is
the amplitude of the motion of one grain (between $0.13 d$ and $0.03 d$) times the number of coherent grains layers. Surprisingly, it is
easy, with only a few synchronized layers of grains, to obtain the high power measured, around $110\, dB$ \cite{H04}. The coherence of the
shocks also explains the seismic wave emitted in the dune \cite{S09}, which can be felt with the feet much further away than the acoustical
sound transmitted through the air \cite{LCCC76}. The essential question is: why would the grains synchronize? It has been proposed recently
that the grains excite a coherent wave within the dune, which in turn vibrates the grains coherently \cite{A04}. But why would the random
collision of the grains create a coherent wave in the first place \cite{S08,S09}? We propose here that the synchronization is due to a
resonance within the flowing layer, as dunes are not needed for sound production. A moving sand grain emits a sound wave during its up and
down motion. If the wave goes to the surface, reflects, and comes back to the grain after exactly one motion period, this sound wave can
slightly push the grain, helping the grain to regularize its motion. It can also induce the grains that are close to move coherently, with the same phase.\\

\begin{figure}[h]
\centering
\includegraphics[width=80mm]{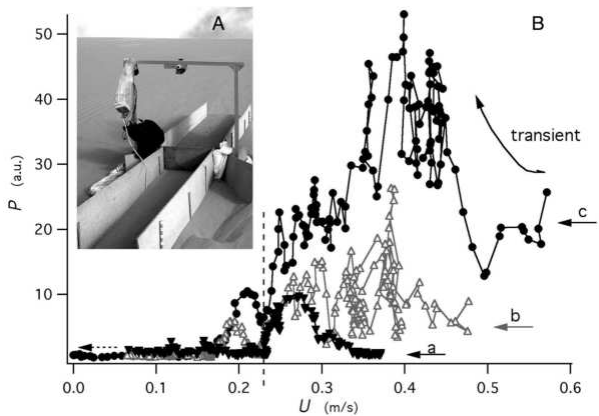}
\caption{\footnotesize{ A: Photo of experiment on dune. A channel, $45\, cm$ wide, $3\, m$ long, was constructed with lateral wood plates,
with a gate in middle, on slip face of singing dune. Using spontaneous flow of grains at their critical angle, two slopes of different
heights are prepared. Then gate is removed, and controlled avalanche is produced). Sound level is recorded simultaneously with the surface
flow. B: Amplitude of sound (pressure level P) as a function of velocity at surface $U$ in middle of channel. Three experiments are shown,
starting with different initial height difference: (a) $5\, cm$, (b) $6.5\, cm$, (c) $10\, cm$. As flow starts (a, b, c arrows), sound takes
some time to develop (transient arrow). With avalanche, height difference decreases, velocity at surface too, eventually stopping (dashed
arrow). Durations of experiments are roughly $15\, s$. In three experiments, sound stops for a surface velocity below $0.23\, m/s$. Bump seen
below (not seen near threshold, curve a), can be ascribed to a secondary sound emission in lower part of channel.}} \label{fig3}
\end{figure}
%

\begin{figure}[h]
\centering
\includegraphics[width=80mm]{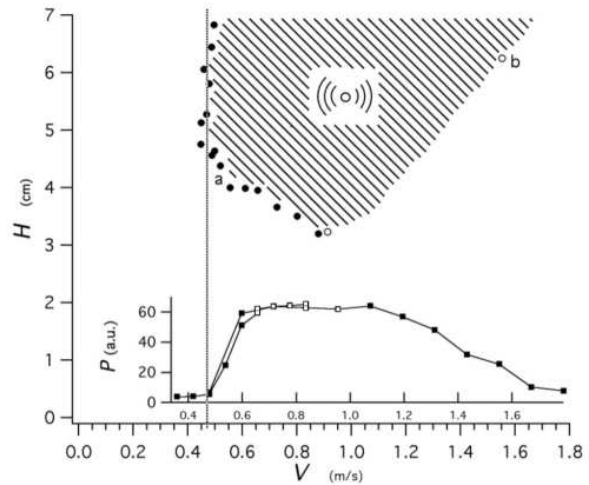}
\caption{\footnotesize{Parameter range for sound emission (shaded area) in laboratory experiment, depending on velocity of blade V and height
of pushed sand $H$. Right threshold occurs when sand is pushed too quickly, so that it is not sheared but projected away (fluidized). Lower
threshold (for $H \sim 3\, cm$) could come from fact that number of pushed grains becomes too small to obtain a proper shear in this
particular shearing geometry (Figure 1B). Left threshold shows that there is no sound emission below a pushing velocity of $0.47\, m/s$.
Inset: variation of sound amplitude P with increasing velocity, from point a to b. Hollow points correspond to measurements where the
microphone saturated.}} \label{fig4}
\end{figure}

Considering the boundary conditions (a fixed sand bed below, and a free moving surface above), this resonance condition corresponds to a
depth of $\lambda/4$, where $\lambda$  is the wavelength (like the resonance of a rod with one end handled and the other one free). For a
smaller depth, no resonance can occur. But for any larger depth, the resonance can occur because the synchronized layers will appear at a
quarter of a wavelength below the free surface. This reasoning is consistent with our field observation that avalanches that are too thin do
not produce any sound, whereas when they become thick enough, even though they may be small in transverse directions, they sing. This height
condition, $H \geq \lambda /4$, can be rewritten as a grain velocity condition. Using $\lambda = c/f$, where $c$ is the sound velocity in the
sheared layer, it gives $fH \geq c/4$. In the flowing layer, the shock frequency $f$ is also the shear rate (constant in depth), so that the
velocity at the surface is $U = fH$. Thus the condition also reads $U \geq c/4$. Controlled experiments have been conducted on the Ghord
Lahmar dune near Foum Agoutir, Tarfaya, Morocco, by constructing a channel in which a controlled avalanche is produced (see
Figure~\ref{fig3}). The results show that there is a threshold surface velocity below which no sound is emitted \cite{S10,S11,S12}. This
threshold gives the sound speed in the flowing layer, $c = 4U = 0.92\, m/s$. The laboratory experiment corroborates the field
experiment. Figure~\ref{fig4} shows the threshold curve in the velocity/depth coordinates. The curve indicates that there is a threshold
velocity (dashed line), below which no sound can be made. Taking into account that the sheared layer is now between the pushed pile and the
static grain bottom, it then gives a resonance condition $H\geq \lambda/2$ (like a string held at both ends), which translates for the pushing
velocity into $V \geq c/2$. This result gives a sound velocity $c = 0.94\, m/s$, which compares well with the one measured in the
field, for the same sand.\\

%


Surprisingly, $0.9\, m/s$ is also the threshold velocity for producing sounds obtained by \cite{B54} when plunging a rod into singing beach
sand. Could the sound emission in this case come from the creation of a supersonic shock wave? A resonance at a quarter of the acoustic
wavelength is well understood in a vertically vibrated layer of grains \cite{RDT90}. The surprising result, here, is the very low sound
velocity: the sound velocity within quartz grains is around $3750\, m/s$ and $330\, m/s$ in air. Considering a mixture of sand and air, a
smaller sound velocity of around $10-33\, m/s$ can be obtained by an effective medium computation \cite{RDT90}, but it is still much larger
than what we observed. Such a low sound velocity, which has also been proposed for singing sand recently \cite{P03}, could come from two
factors. First this sound velocity is observed in the sheared layer. As the sound passes in part through the contacts between the grains, the
reduced number of contacts in flowing sand makes it more difficult for the pressure wave to propagate. The second reason can be related to
the fact that not all sand is musical, and not all well-sorted dunes sing. The musical property probably comes from a special surface state
of the singing sand grains \cite{JYW98}. The ability to sing has previously been ascribed to the presence of a silica gel layer on the grain
surface \cite{GLK97}, known as desert glaze \cite{CWG93,PT90}. The importance of such a layer is shown by the fact that after intensive use
the grains lose their sound-emitting properties. The surface state of the grains seems to reduce the sound velocity from its normal value to
a much lower one. This could explain why the threshold (in height or velocity) for sound emission is not constant across geographical
locations. In China, several people pushing simultaneously as much sand as possible are needed to produce sound. It could also explain why
avalanches never produce any sound normally, as it is difficult to produce a flowing layer that would be thick enough. This surprisingly low
sound velocity for musical sand, and its sensitivity in other parameters, such as humidity, temperature \cite{C23}, or dust
\cite{H86}, should now warrant particular attention.\\


This age-old geological mystery of singing avalanches reveals an original way of producing sound. The sound comes from the synchronized
motion of grains, and it is shown here that they synchronize because of a resonance inside the sheared layer. In this way, singing avalanches
may be understood as a new type of instrument, as the frequency is not controlled by the resonance, but imposed by the motion of the grains.
If a resonance is still needed, it is not to select the frequency, but to produce the necessary self-synchronization of the grains.\\

\begin{acknowledgments}
Thanks are due to Herv\'{e} Belot, who participated in the first experiments on the dunes, Marc Elsen, who was the first to train on the
singing sand in the lab, Margherita Peliti, who trained on a first version of the laboratory experiment, Laurent Quartier for constructing
the laboratory experiments, Bruno Andreotti for enforcing the idea that the sound could be emitted in the air as a loudspeaker, and
Andr\'{e}s Illane-Campo, for guiding us in Copiapo to the singing dunes. This work was possible only thanks to an Action Concert\'{e}e
Incitative Jeune Chercheur.\\
\end{acknowledgments}

\end{document}